%% file: main.tex
\pdfoutput=1
\documentclass[sigconf,screen]{acmart}

\AtBeginDocument{%
  }

\usepackage[ruled,linesnumbered,vlined]{algorithm2e}
\usepackage{pifont} 
\usepackage{multirow}
\usepackage{graphicx}
\usepackage{color}
\usepackage[utf8]{inputenc}
\usepackage{textcomp}
\usepackage{soul}
\usepackage{subfigure}
\usepackage{makecell}
\usepackage{listings}
\usepackage{amssymb}
\usepackage{enumitem}
\usepackage{fancybox}
\usepackage{verbatim}
\usepackage[utf8]{inputenc} 
\definecolor{mylightblue}{RGB}{213, 235, 254}
\definecolor{mylightpink}{RGB}{245, 216, 235}
\newcommand{\highlightpink}[1]{\sethlcolor{mylightpink}\hl{#1}}
\newcommand{\highlightblue}[1]{\sethlcolor{mylightblue}\hl{#1}}

\copyrightyear{2026}
\acmYear{2026}
\setcopyright{rightsretained}
\acmConference[ICSE '26]{2026 IEEE/ACM 48th International Conference on Software Engineering}{April 12--18, 2026}{Rio de Janeiro, Brazil}
\acmBooktitle{2026 IEEE/ACM 48th International Conference on Software Engineering (ICSE '26), April 12--18, 2026, Rio de Janeiro, Brazil}
\acmPrice{}
\acmDOI{10.1145/3744916.3764530}
\acmISBN{979-8-4007-2025-3/26/04}

\newboolean{showcomments}
\setboolean{showcomments}{true}
\ifthenelse{\boolean{showcomments}}
 { \newcommand{\mynote}[2]{
      \fbox{\bfseries\sffamily\scriptsize#1}
        {\small$\blacktriangleright$\textsf{\emph{#2}}$\blacktriangleleft$}}}
        { \newcommand{\mynote}[2]{}}

\definecolor{ForestGreen}{RGB}{0,180,0}
\definecolor{ForestGreen}{RGB}{0,180,0}

\newcommand{\hhre}[1]{{#1}}

\newcommand{\eg}{\emph{e.g.,}}
\newcommand{\ie}{\emph{i.e.,}}
\newcommand{\etal}{\emph{et al.}}

\definecolor{lightergray}{rgb}{0.9,0.9,0.9} % 浅灰
\newcommand{\tool}{\textsc{EasyLink}}

\begin{document}


% \title{Something something simple over hard: Issue-commit linking with vector DB and LLM}
% \title{Issue-Commit Linking with Vector DB and LLM: A Simple Yet Effective Approach}
% \title{EasyLink: Issue-Commit Linking with Vector DB and LLM for a Simple Yet Effective Approach}
% \title{EasyLink: Simple Over Hard for Issue-Commit Linking with Vector DB and LLM}
% \title{EasyLink: Simple Over Hard for Issue-Commit Linking with Vector Search and LLM-assisted Reranking}
\title[Rethinking Issue-Commit Linking with LLM-Assisted Retrieval]{Back to the Basics: \\ Rethinking Issue-Commit Linking with LLM-Assisted Retrieval}

% \author{Huihui Huang\textsuperscript{$\diamondsuit$}, Ratnadira WIDYASARI\textsuperscript{$\diamondsuit$}, Ting Zhang\textsuperscript{$\diamondsuit$}, Ivana Clairine IRSAN\textsuperscript{$\diamondsuit$}, Jieke Shi\textsuperscript{$\diamondsuit$}, Han Wei Ang\textsuperscript{$\spadesuit$}, Frank Liauw\textsuperscript{$\spadesuit$}, Eng Lieh Ouh\textsuperscript{$\diamondsuit$}, Lwin Khin Shar \textsuperscript{$\diamondsuit$}, Hong Jin Kang\textsuperscript{$\clubsuit$}, and David Lo\textsuperscript{$\diamondsuit$}}

% \thanks{$^\dagger$Jieke Shi is the corresponding author.}

% \affiliation{%
%   \institution{\textsuperscript{$\diamondsuit$}School of Computing and Information Systems, Singapore Management University, Singapore}\country{}
% }
% \affiliation{%
%   \institution{\textsuperscript{$\spadesuit$}GovTech, Singapore}\country{}
% }
% \affiliation{%
%   \institution{\textsuperscript{$\clubsuit $}School of Computer Science, University of Sydney, Australia}\country{}
% }

\author{Huihui Huang\textsuperscript{$\diamondsuit$}, Ratnadira Widyasari\textsuperscript{$\diamondsuit$}, Ting Zhang\textsuperscript{$\heartsuit$}, Ivana Clairine Irsan\textsuperscript{$\diamondsuit$}, Jieke Shi\textsuperscript{$\diamondsuit$}, Han Wei Ang\textsuperscript{$\spadesuit$}, Frank Liauw\textsuperscript{$\spadesuit$}, Eng Lieh Ouh\textsuperscript{$\diamondsuit$}, Lwin Khin Shar \textsuperscript{$\diamondsuit$}, Hong Jin Kang\textsuperscript{$\clubsuit$}, and David Lo\textsuperscript{$\diamondsuit$}}

\thanks{$^\dagger$Jieke Shi and Ting Zhang are the corresponding authors.}

\affiliation{%
  \institution{\textsuperscript{$\diamondsuit$}School of Computing and Information Systems, Singapore Management University, Singapore}\country{}
}
\affiliation{%
  \institution{\textsuperscript{$\heartsuit$}Faculty of Information Technology, Monash University, Australia}\country{}
}
\affiliation{%
  \institution{\textsuperscript{$\spadesuit$}GovTech, Singapore}\country{}
}
\affiliation{%
  \institution{\textsuperscript{$\clubsuit$}School of Computer Science, University of Sydney, Australia}\country{}
}

\affiliation{%
  \institution{\{hhhuang, ratnadiraw.2020, ivanairsan, jiekeshi, elouh, lkshar, davidlo\}@smu.edu.sg \\ ting.zhang@monash.edu,
  \{ang\_han\_wei, frank\_liauw\}@tech.gov.sg, hongjin.kang@sydney.edu.au}\country{}
}

\renewcommand{\shortauthors}{Huang et al.}

\begin{abstract}

Issue-commit linking, which connects issues with commits that fix them, is crucial for software maintenance. Existing approaches have shown promise in automatically recovering these links. Evaluations of these techniques assess their ability to identify genuine links from plausible but false links. However, these evaluations overlook the fact that, in reality, when a repository has more commits, the presence of more plausible yet unrelated commits may interfere with the tool in differentiating the correct fix commits. To address this, we propose the Realistic Distribution Setting (RDS) and use it to construct a more realistic evaluation dataset that includes 20 open-source projects. By evaluating tools on this dataset, we observe that the performance of the state-of-the-art deep learning-based approach drops by more than half, while the traditional Information Retrieval method, VSM, outperforms it.

Inspired by these observations, we propose \textbf{\tool{}}, 
which utilizes a vector database as a modern Information Retrieval technique.
To address the long-standing problem of the semantic gap between issues and commits, \tool{} leverages a large language model to rerank the commits retrieved from the database. 
Under our evaluation, \tool{} achieves an average Precision@1 of 75.03\%, improving over the state-of-the-art by over four times. Additionally, this paper provides practical guidelines for advancing research in issue-commit link recovery.

\end{abstract}

\begin{CCSXML}
<ccs2012>
   <concept>
       <concept_id>10011007.10011006.10011073</concept_id>
       <concept_desc>Software and its engineering~Software maintenance tools</concept_desc>
       <concept_significance>500</concept_significance>
       </concept>
   <concept>
       <concept_id>10010147.10010178</concept_id>
       <concept_desc>Computing methodologies~Artificial intelligence</concept_desc>
       <concept_significance>500</concept_significance>
       </concept>
</ccs2012>
\end{CCSXML}

% \ccsdesc[500]{Software and its engineering~Software maintenance tools}
% \ccsdesc[500]{Computing methodologies~Artificial intelligence}


\keywords{Issue-Commit Link Recovery, Software Traceability}

% \received{30 November 2024}

% \setcopyright{none} % to remove the copyright notice
% \settopmatter{printacmref=false} % to remove the ACM Reference Format
% \renewcommand\footnotetextcopyrightpermission[1]{}

\maketitle

\input{1_introduction}
\input{2_background}

\input{4_dataset_construction}

\input{3_simple_method}

\input{5_experimental_setup}

\input{5_results}
\input{6_discussion}

\input{8_threats}

\input{7_related_work}

\input{9_conclusion}
\input{acknowledgement}

% \section{Introduction}

% this is the introduction

% \hj{Huihui, can we try to balance the columns for the references?  I think there's a latex to include, and then \\balance  should work }
\balance
\bibliographystyle{ACM-Reference-Format}
\bibliography{ref}

% \appendix

\end{document}

%% file: 1_introduction.tex
\section{Introduction}
\label{sec:introduction}

Software traceability involves establishing relationships between different software artifacts and is essential for safety-critical systems~\cite{cleland2012software, 10.1145/2593882.2593891}. 
A critical task in this domain is {\it issue-commit linking}, which connects issues, \ie{} bug reports, to the commits that resolve them~\cite{kondo2022empirical}, playing a vital role in software provenance.
It also plays a key role for developers in assessing security risks~\cite{nguyen2022hermes,nguyen2022vulcurator,li2024patchfinder} and gaining deeper insights about security flaws~\cite{wu2025commitshield,meneely2013patch}.

Prior studies~\cite{bachmann2009software,10.1145/2641580.2641592} have revealed that many issue-commit links can be missing during the development of large-scale projects. Manually recovering these links is not only time-consuming but error-prone, even for experienced developers~\cite{ruan2019deeplink}.

To address this, several studies~\cite{zhang2023ealink,lin2021traceability,dong2022semi,ruan2019deeplink,sun2017improving,rath2018traceability} have proposed learning-based approaches to automatically recover issue-commit links, achieving strong performance on datasets collected from open-source repositories.

The evaluation method requires the tool to distinguish fix commits (\ie{} ``true links'') from plausible but non-fix commits (\ie{} ``false links''). 
However, prior studies' evaluations often lack realism. 
\ding{182} Some studies use an unrealistic time window to select commits. 
For example, studies~\cite{ruan2019deeplink,nguyen2012multi,sun2017frlink,bachmann2010missing} rely on a narrow 7-day time window to select potential fix commits, which may miss many true links. Another example involves studies~\cite{dong2022semi,rath2018traceability} that require the issue close time to select commits, but in practice, when commits are missing, the issue may not have a close time.
\ding{183} Another limitation is the unrealistic false link distributions in prior evaluation datasets.
Some~\cite{lin2021traceability,ruan2019deeplink} use evaluation datasets where the number of false and true links is equal, an unrealistic assumption since false links far outnumber true links in reality~\cite{dong2022semi}.
Recently, Zhang \etal{}~\cite{zhang2023ealink} addressed this issue by using an imbalanced dataset with a fixed number of false links per issue. 
Still, this method overlooks a crucial factor: a higher commit frequency results in a larger pool of plausible commits for the tool to differentiate from the actual fix commit,
making the task inherently more difficult.

Therefore, we propose to construct a more realistic evaluation dataset under the \textbf{Realistic Distribution Setting} (RDS).
We began with an in-depth analysis of the issues and commits from open-source repositories used in prior studies~\cite{zhang2023ealink, dong2022semi, lin2021traceability, mazrae2021automated}.
According to our observation, after fetching all commits in each repository, approximately 97\% of fix commits were made within one year from issue creation. \hhre{These findings suggest that, in practice, the corresponding fix commit for an issue is likely to be found among the commits made within one year after the issue's creation.} 
We thus include the genuinely linked commit as the true link and consider all non-fix commits made within one year of an issue's creation as candidate commits linked to the given issue as false links when constructing evaluation datasets.

After collecting issues and commits from 20 open-source projects analyzed in prior studies~\cite{zhang2023ealink, dong2022semi, lin2021traceability, mazrae2021automated}, 
we first successfully replicate the performance of the state-of-the-art method, EALink~\cite{zhang2023ealink}, using
the original evaluation method they reported.
However, when we switched the experimental setup to the Realistic Distribution Setting, we find that the average Precision@1 of EALink dropped to 14.43\%, surprisingly underperforming the Vector Space Model (VSM)~\cite{antoniol2002recovering}, a traditional Information Retrieval (IR) technique, which achieves a Precision@1 of 46.59\%, suggesting that there is great room for improving issue-commit linking using IR techniques. Meanwhile, as recent years have witnessed huge advances in the IR field, modern IR techniques that use embedding models to capture semantic-level similarities have demonstrated better performance than those traditional techniques evaluated in previous studies~\cite{dong2022semi,zhang2023ealink,lin2021traceability} that measure token-level textual similarities, such as VSM~\cite{antoniol2002recovering} and Latent Semantic Indexing (LSI)~\cite{marcus2005recovery}. This also presents an opportunity to boost issue-commit linking performance and motivates our paper.

In this paper, we propose \textbf{\tool{}}, 
a novel method that integrates modern IR techniques and Large Language Models (LLMs) to enhance issue-commit linking performance in realistic settings by capturing deeper semantic relationships between issues and commits.
\tool{} operates in two stages. The first stage fetches a set of commits that share similarities to a given issue, and the second stage reranks them by their relevance to the issue. 
Concretely, in the first stage, \tool{} leverages recent advances in Information Retrieval by using a modern vector database (optimized for high-dimensional similarity search)~\cite{douze2024faiss} to fetch the most similar commits for each issue. 
% This stage provides an initial list of results that will be refined in the second stage. 
In the second stage, \tool{} prompts an LLM (GPT-4o~\cite{gpt-4o}) to rerank the results. We use LLMs for their strong ability to capture the semantic relationship between issues and commits, bridging the semantic gap, as the most similar commit does not guarantee that it is the fix for the issue~\cite{zhang2023ealink, lin2021traceability, dong2022semi}.
\tool{} achieves an average Precision@1 of 75.03\%, outperforming EALink~\cite{zhang2023ealink} by four times in our realistic datasets and by more than 30\% in Precision@1 on EALink's original evaluation setup.
We also conduct an ablation analysis on each of \tool{}'s two stages.
Our results demonstrate that advances in IR, such as context-aware dense embeddings~\cite{wang2020minilm,song2020mpnet,neelakantan2022text} and efficient vector search~\cite{douze2024faiss,wang2021milvus}, have been largely overlooked in the software traceability literature. Notably, even the out-of-the-box use of a vector database achieves a high average Precision@1 of 61.57\%.
Additionally, the strong performance of the reranking step, which improves Precision@1 by 13.46\%, highlights the capability of LLMs to bridge the semantic gap between issues and commits~\cite{rath2018traceability,guo2017semantically}.
Finally, we discuss the lessons learned from our work, such as the need to consider modern IR baselines, for future research on software traceability. Our implementation has been made available at~\cite{replicationPackage}.

This paper makes the following contributions:

\begin{enumerate}[leftmargin=*, itemsep=2pt]

    \item \textbf{Constructing a more realistic evaluation dataset:} To achieve a more realistic evaluation, we propose the Realistic Distribution Setting (RDS), which adjusts the number of candidate commits for generating false links according to the quantity of commits in the repository. This results in a dataset that more accurately reflects real-world practices. 
    \item \textbf{A comprehensive evaluation benchmark:} We include the datasets from recent studies~\cite{zhang2023ealink, dong2022semi, lin2021traceability, mazrae2021automated}. 
    Our benchmark includes 9,319 issues from 20 projects, with an average of 1,530 false links constructed per issue. To the best of our knowledge, this dataset is the largest in the literature.
    
    \item \textbf{Reevaluation of the state-of-the-art approach:} 
    After successfully replicating the strong performance of EALink~\cite{zhang2023ealink}, the state-of-the-art approach,
    we switched the evaluation procedure under RDS, which offers a greater challenge due to a higher number of false links.
    On the same set of projects used in the evaluation of EALink, its average Precision@1 of 53.67\%
    decreases to 28.05\%,
    underperforming traditional IR baselines.

\item \textbf{A new state-of-the-art for issue-commit linking:} We propose \tool{}, which leverages modern IR techniques, including an off-the-shelf database FAISS~\cite{douze2024faiss}, and addresses the problem of semantic gap~\cite{rath2018traceability,guo2017semantically} by prompting an LLM to rerank the retrieved results. 
    In our evaluation, \tool{} improves over EALink in average Precision@1 from 14.43\% to 75.03\%.

\end{enumerate}

The rest of the paper is organized as follows.~\autoref{sec:background} outlines the background of this work, including the limitations of existing work.~\autoref{sec:dataset_construction} details the evaluation dataset construction under the Realistic Distribution Setting (RDS).~\autoref{sec:simple_method} introduces \tool{}.~\autoref{sec:experimental_setup} describes the experimental setup.~\autoref{sec:results} presents the experimental results.~\autoref{sec:discussion} revisits the classical baseline, discusses the lessons learned and threats to validity. ~\autoref{sec:related_work} reviews related work.
Finally, ~\autoref{sec:conclusion} concludes the paper.

%% file: 2_background.tex
\section{Background}
\label{sec:background}

\subsection{Issue-Commit Linking Recovery}

\definecolor{lightergray}{rgb}{0.95, 0.95, 0.95}

% 紧凑并自动换行的 listings 全局设置
\lstset{
  basicstyle=\ttfamily\footnotesize,
  breaklines=true,
  aboveskip=0pt,
  belowskip=0pt,
  xleftmargin=0.6em,
  xrightmargin=0.6em,
  columns=fullflexible,
  keepspaces=true
}

\begin{figure}[!t]
\centering

\doublebox{%
  \begin{minipage}{0.95\linewidth} % 与当前栏宽一致

    % ===== 黑色标题条（宽度 = \linewidth - 2\fboxsep）=====
    {\setlength{\fboxsep}{6pt}%
    \colorbox{black}{%
      \parbox{\dimexpr\linewidth-2\fboxsep\relax}{%
        \color{white}\bfseries
        An example of an issue, the incorrectly top-ranked commit, and the correct commit.%
      }%
    }}%

    \vspace{0.4em}

    % ===== 第一段灰底说明（同样精确扣除 2\fboxsep）=====
    {\setlength{\fboxsep}{6pt}%
    \colorbox{lightergray}{%
      \parbox{\dimexpr\linewidth-2\fboxsep\relax}{%
        \noindent The input issue, with summary and description, is shown below.\\
        \small Issue ID: NETBEANS-803
      }%
    }}

    {\setlength{\fboxsep}{6pt}%
    \colorbox{white}{%
      \parbox{\dimexpr\linewidth-2\fboxsep\relax}{%
      \ttfamily\footnotesize 
Summary:\\
nb-javac 11 upgrade in NetBeans\\
Description:\\
Should cover below tasks in NetBeans - nb-javac 11 testing :\\
\textbullet\ Run tests for modules java.completion, java.editor, \\ 
\hspace*{1em}java.editor.base, java.hints, java.source,\\ 
\hspace*{1em}java.source.base, lib.nbjavac \\
\textbullet\ Update libs.javacimpl and libs.javacapi jars, upload updated \\
\hspace*{1em}nb-javac jars\\
\textbullet\ Upload nb-javac module jars in update center
      }%
    }}

    % ===== 代码块 1（不放在 colorbox 里，避免冲突）=====
% \begin{lstlisting}
% Summary:
% nb-javac 11 upgrade in NetBeans
% Description:
% Should cover below tasks in NetBeans - nb-javac 11 testing:
% Run tests for modules java.completion, java.editor, 
% java.editor.base, java.hints, java.source, 
% java.source.base, lib.nbjavac
% Update libs.javacimpl and libs.javacapi jars, 
% upload updated nb-javac jars
% Upload nb-javac module jars in update center
% \end{lstlisting}

    \noindent\rule{\linewidth}{0.6pt}

    % ===== 第二段灰底说明 =====
    {\setlength{\fboxsep}{6pt}%
    \colorbox{lightergray}{%
      \parbox{\dimexpr\linewidth-2\fboxsep\relax}{%
        \noindent The top-ranked commit message in the initial results is shown below but is \textbf{incorrect}.\\
        \small Commit ID: 4fd115aeae3b8423de9ced22d52914e60a1c5800.
      }%
    }}

    {\setlength{\fboxsep}{6pt}%
    \colorbox{white}{%
      \parbox{\dimexpr\linewidth-2\fboxsep\relax}{%
      \ttfamily\footnotesize 
Updation for external nb-javac jar in libs.javacapi and \\
libs.javaimpl modules with nb-javac jar for jdk-12
      }%
    }}

% \begin{lstlisting}
% Updation for external nb-javac jar in libs.javacapi and
% libs.javaimpl modules with nb-javac jar for jdk-12
% \end{lstlisting}

    \noindent\rule{\linewidth}{0.6pt}

    % ===== 第三段灰底说明 =====
    {\setlength{\fboxsep}{6pt}%
    \colorbox{lightergray}{%
      \parbox{\dimexpr\linewidth-2\fboxsep\relax}{%
        \noindent The \textbf{correct} commit message is shown below.\\
        \small Commit ID: 3055661e4dd1c7d587012c917cbec31b27ae9e34
      }%
    }}

    {\setlength{\fboxsep}{6pt}%
    \colorbox{white}{%
      \parbox{\dimexpr\linewidth-2\fboxsep\relax}{%
      \ttfamily\footnotesize 
Uptake nb-javac 11 jars for java tests runtime
      }%
    }}

% \begin{lstlisting}
% Uptake nb-javac 11 jars for java tests runtime
% \end{lstlisting}

  \end{minipage}%
}

\caption{Example of an issue with the incorrect top-ranked commit and the correct commit. An issue-commit linking approach has to bridge the semantic gap and distinguish the correct commit from similar ones in a potentially large set.}
\label{example}
\end{figure}

Issue-commit links play an essential role in maintaining software traceability, supporting critical tasks such as impact analysis~\cite{aung2020literature,hindle2008large,purushothaman2005toward}, regression testing~\cite{naslavsky2007using}, and project management~\cite{panis2010successful}. 
Due to the high cost of manually maintaining these links, they are often incomplete~\cite{izadi2022linkformer}, highlighting the need for automated methods.
Automatic issue-commit linking considers a large set of commits that are made in the period after each issue's creation and identifies the right commits that address the issue.~\autoref{example} shows an example that demonstrates the challenge of automatically linking issues to commits. 
It presents an issue along with two commits. 
The first commit shares more matched words and has the highest similarity to the issue, but it is not the fix, while the second commit is the correct fix commit.
An issue-commit linker that considers keyword counts would incorrectly prioritize the first commit as there are a greater number of matches of the keyword \texttt{nb-javac}. 
While both commits mention \texttt{nb-javac}, the top-ranked commit updates the JAR for JDK-12, whereas the issue specifies that nb-javac 11 should be upgraded, tested, and deployed. 
This highlights two challenges.
First, it shows the need for approaches that go beyond surface-level similarities.
Second, it demonstrates the sensitivity of the evaluations of their effectiveness to the number of commits that share keywords or resemble the ground-truth commit.

Many automatic issue-commit link recovery methods have been proposed. 
Some studies employ traditional feature- and rule-based methods ~\cite{schermann2015discovering,wu2011relink,nguyen2012multi,bachmann2010missing}, which rely on predefined heuristics such as keyword matching or the recency of the commits. 
However, these heuristics tend to be inadequate
~\cite{bird2009fair}.
Some approaches~\cite{mazrae2021automated,rath2018traceability,sun2017improving,le2015rclinker,sun2017frlink} adopt traditional machine learning techniques, such as support vector machines, to reduce the reliance on manual rules. 
Recently, deep learning methods ~\cite{ruan2019deeplink,xie2019deeplink,guo2017semantically} have demonstrated improved performance. 
A key challenge of the task is the semantic gap that exists between issues and commits~\cite{zhang2023ealink,lin2021traceability,dong2022semi}.
Prior work~\cite{lin2021traceability} attempted to address this issue using BERT-based methods, such as CodeBERT~\cite{feng2020codebert}, which leverage contextual understanding.
The state-of-the-art method, EALink~\cite{zhang2023ealink}, employs knowledge distillation to transfer knowledge to a smaller model and utilizes multi-task learning to improve both accuracy and efficiency.

\subsection{Limitations in Evaluation}

The experiments in prior studies evaluate approaches based on their ability to distinguish the true
link from false links constructed using other commits. However, these studies use unrealistic methods to construct false links and the evaluation dataset. 

\textbf{Unrealistic Time Window Selection}:
Some works~\cite{ruan2019deeplink,nguyen2012multi,sun2017frlink,bachmann2010missing} assume that fix commits fall within a 7-day time window before or after the issue’s create/update/close time or the comment create time, treating all other commits in this period as non-fix commits.
However, this approach is inadequate when fixes take a longer time, requiring tools to distinguish the correct fix commit from more plausible commits.
From our preliminary analysis of selected projects, we found that only 59\% of issues had a corresponding fix commit within seven days of their creation.
Additionally, some works~\cite{dong2022semi,rath2018traceability} require the issue’s close time to select commits. 
Under a practical scenario, the issue fix/close date is unknown—precisely also why issue-commit recovery is necessary—and approaches need to distinguish the fix commit while the issue remains open.
Therefore, selecting commits based on the issue's fix/close date may also lack realism.

\begin{figure}[t!]
    \centering
    \includegraphics[width=\linewidth]{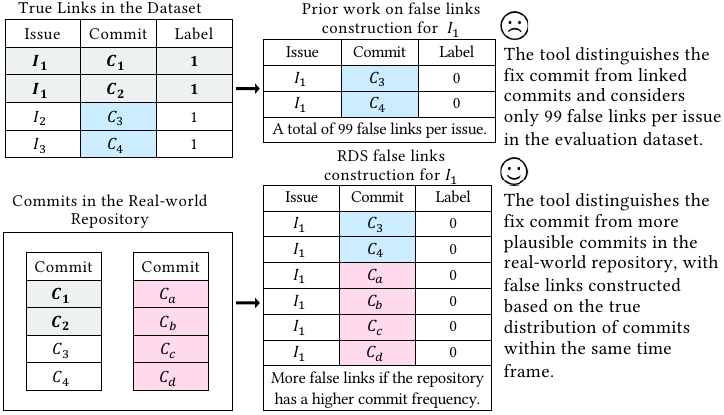}
   \caption{Illustration of the evaluation limitation in prior work. $C_1$ and $C_2$ are the fix commits of issue $I_1$. False links will be constructed for $I_1$.
\highlightblue{$C_3$} and \highlightblue{$C_4$} are commits already linked to the issue in the true links dataset.  
\highlightpink{$C_a$}, \highlightpink{$C_b$}, \highlightpink{$C_c$}, and \highlightpink{$C_d$} are commits present in the repository but not in the true links dataset, and they are ensured not to be the fix commit of $I_1$. 
   }
    \label{fig:example_false_link_gen}

\end{figure}

\textbf{Unrealistic False Link Distribution:} When constructing the evaluation dataset, some works~\cite{ruan2019deeplink,lin2021traceability,mazrae2021automated} use a balanced dataset (\ie{} an equal number of false links and true links) to evaluate their tools. 
This evaluation setting is unrealistic because, in reality, false links outnumber true links~\cite{dong2022semi}.
Zhang \etal{}~\cite{zhang2023ealink} address the balanced dataset limitation by constructing a fixed number of 99 false links per issue. 
While this results in more false links than true links, it still overlooks that the number of potentially linked commits depends on the number of commits made in the same time period as the ground-truth link. 
Moreover, as shown in \autoref{fig:example_false_link_gen}, when sampling unrelated commits to construct the false links, prior work~\cite{zhang2023ealink} only selects commits from other ground-truth links. 
In other words, they construct false links for an issue by connecting it only to other commits that are already linked to other issues.
In practice, an approach has to consider all commits from the same time period, regardless of whether they are linked to specific issues.
This is a superset of commits compared to the ground-truth links. 
In~\autoref{sec:dataset_construction}, we show that the number of commits made within a one-year time frame is at least 80\% larger than the constant number of 99 commits considered by Zhang \etal{}~\cite{zhang2023duplicate}.
As a result, the evaluation setups of prior work may not adequately reflect how these approaches would be used in practice.

%% file: 4_dataset_construction.tex
\section{Evaluation Dataset Construction Under the Realistic Distribution Setting}
\label{sec:dataset_construction}

This section describes our new evaluation setting, \textbf{Realistic Distribution Setting} (RDS), for constructing a more realistic evaluation dataset. 
By adaptively constructing false links for an issue based on the commits in the period after each issue creation, aligning with the repository's development activity, our method addresses the limitations of prior evaluations.

We combine datasets from four recent studies~\cite{zhang2023ealink, lin2021traceability, dong2022semi, mazrae2021automated} into a unified benchmark covering 20 projects, ensuring broader coverage and a consistent benchmark for comparison. 
Evaluations in the previous studies~\cite{zhang2023ealink, lin2021traceability, mazrae2021automated} did not use a shared benchmark, and each of their datasets had only up to 12 projects~\cite{mazrae2021automated}. 
% without a shared benchmark
Our dataset fills the need for a large, shared benchmark.

\subsection{Ground Truth Dataset Selection}\label{sec:ground-truth}
We selected datasets from four recent studies, covering 20 open-source software projects over a span of 20 years. These datasets use explicit issue tags (e.g., ``\#123'', ``JIRA-456'') in commit messages to construct the true links.
We include the same issues and true links from these datasets, resulting in a total of 9,319 issues in the benchmark. 
For datasets with incomplete information detected through a manual check, we re-fetched the necessary data, such as issue comments, code diffs, and committed files (\ie{} the source code files after applying the commits).

\begin{itemize}[leftmargin=*]
    \item \textbf{From Zhang \etal{}'s dataset~\cite{zhang2023ealink}}, we include the issues from all six projects, \textit{Ambari}, \textit{Calcite}, \textit{Groovy}, \textit{Ignite}, \textit{Isis}, and \textit{Netbeans}, in our benchmark. 
    
    \item \textbf{From Dong \etal{}'s dataset~\cite{dong2022semi}}, we include the issues from five projects, \textit{Pig}, \textit{Maven}, \textit{Infinispan}, \textit{Drools}, and \textit{Derby}.
    This dataset also includes \textit{Groovy}, which is already included in our benchmark from Zhang \etal{}'s dataset~\cite{zhang2023ealink}.
    As they were missing issue comments, code diffs, and committed files, we refetched these data.

    \item \textbf{From Lin \etal{}'s dataset~\cite{lin2021traceability}}, we include the issues from all three projects: \textit{Pgcli}, \textit{Flask}, and \textit{Keras}. As the issue comments and committed files were missing, we refetched them.
    
    \item \textbf{From Mazrae \etal{}'s dataset~\cite{mazrae2021automated}}, 
    we include the issues from six projects, \textit{Beam}, \textit{Flink}, \textit{Freemarker}, \textit{Airflow}, \textit{Arrow}, and \textit{Cassandra}, out of 12 projects. 
    The six other projects from this dataset are already included in Zhang \etal{}'s dataset~\cite{zhang2023ealink}. %\hhre{
    As this dataset provides only postprocessed data, we re-fetched the original issue summaries, issue descriptions, issue comments, commit messages, code diffs, and committed files.
    %}
\end{itemize}

\subsection{Preparing Commits}\label{sec:commits-pool}

For constructing false links, the set of all commits in each repository is required.
First, we clone the GitHub repository locally, which enables faster access to commit data without API requests. 
From the cloned repository, all commit IDs are obtained using the \texttt{git log} command. 
Each commit is then processed to retrieve its metadata, including the parent commit IDs, author, committer, commit time, and commit message, using \texttt{git show}. Additionally, the list of modified file paths for each commit is identified, and their content at the specific commit state is retrieved. 
The corresponding code diffs are also extracted and stored.~\autoref{tab:merged_table} presents the number of commits fetched for each project in the column ``\#Commits''.

The following information is fetched during the process:

\begin{itemize}[leftmargin=*]
    \item \textbf{Commit ID}: A unique hash value assigned to each commit, serving as its identifier.
    \item \textbf{Parent Commit IDs}: The hash(es) of the immediate predecessor commit(s) of the current commit.
    \item \textbf{Author}: The person who originally wrote the changes.
    \item \textbf{Committer}: The person who applied the changes to the repository.
     \item \textbf{Commit Time}: The commit's creation timestamp.%The timestamp when the commit was made.
    \item \textbf{Commit Message}: The description of the commit.
    \item \textbf{Changed Files}: List of paths of the files modified in the commit.
    \item \textbf{Code Diffs}: The %specific 
    changes introduced by the commit.
    \item \textbf{Committed Files}: Source code files after applying the commit.
\end{itemize}
Note that while our work only requires the commit ID, commit time, and commit message, we collect all commit information since other approaches may utilize this additional data.

\begin{figure*}[t!]
    \centering
    \includegraphics[width=\linewidth]{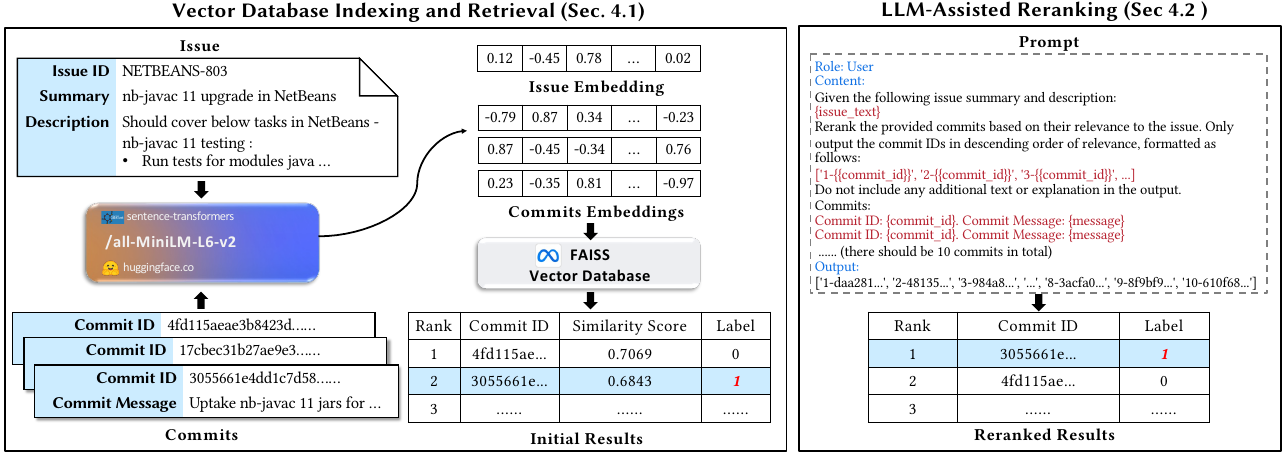}
   \caption{Overview of \tool{}. \tool{} consists of two key steps—the first step utilizes a vector database to retrieve initial ranked results, and the second step prompts an LLM to rerank the results. 
   % \hj{The section numbers should be checked and updated before submission}\hh{sure}
   }
    \label{fig:method}
\end{figure*}

\subsection{Constructing False Links}\label{sec:eval_method}
\input{4_alg}

For a more realistic evaluation dataset, the number of false links should match the actual number of commits that may be viable for linking to each issue in practice. 
For each issue in the evaluation dataset, under the Realistic Distribution Setting (RDS), false links will be adaptively constructed to better align with the repository's level of activity.
% Rather than considering all repository commits as candidates, an impractical approach, w
We include all commits submitted in the time window within which the true commit can appear. 
To determine the size of the time window, we perform an analysis and find that approximately 97\% of all issues have their ground truth commit submitted within one year of their creation (7-day time window: 59\%, 30-day time window: 77\%, 6-month time window: 92\%).
This observation aligns with previous studies analyzing bug reports (\eg{} Rodrigues \etal{}~\cite{rodrigues2020soft}, Zhang \etal{}~\cite{zhang2023duplicate}). 
Consequently, we apply a one-year time window to select candidate commits from the commits pool (as described in~\autoref{sec:commits-pool}) for each issue. 
Specifically, for a true link denoted as \( t_i = \{I_i, C_i\} \), where \( I_i \) represents the issue and \( C_i \) the commit in the true link, the candidate commit \( C_j \) is determined using~\autoref{equ:candidate_commit}. If the commit time of \( C_j \) is within one year after the creation time of \( I_i \), we treat \( C_j \) as a candidate commit. By linking \( C_j \) to \( I_i \), we generate the false link \( f_i = \{I_i, C_j\} \).

\begin{equation}
\label{equ:candidate_commit}
\begin{aligned}
\text{is\_candidate}(I_i, C_j) &= created(I_i) \leq committed(C_j) \\
                           &\land\; committed(C_j) \leq created(I_i) + \epsilon, \\
\epsilon &= \text{one year.}
\end{aligned}
\end{equation}

\autoref{alg:false_link_gen} details the construction of the evaluation dataset.
We first split the entire ground-truth dataset into training and test sets following a 4:1 ratio~\cite{zhang2023ealink,dong2022semi,ruan2019deeplink}.
% \zt{maybe we should justify the ratio}\hh{i added some paper in the citation which use this ratio also.} 
The inputs to the algorithm are the true links from the test set and the commits pool for the project, which is prepared using the process described in~\autoref{sec:commits-pool}. 
First, following the method used by EALink~\cite{zhang2023ealink}, we randomly sample up to 1,000 unique issue IDs from the test set, forming the list $\mathit{sampled\_issue\_id\_list}$. If fewer than 1,000 unique issues are available, all issues are included. Then, for each $\mathit{issue\_id}$ in this list, we extract the true links for that issue from the test set. 
We use the term ``links'' (plural) because an issue may be linked to more than one commit~\cite{zhang2023ealink}. 
In line 6 of the algorithm, we extract only the issue information for the issue being processed (including issue ID, summary, description, etc.), denoted as $\mathit{issue}$. 
% Next, as shown in lines 7 to 10 of the algorithm, we filter the $\mathit{candidate\_commits}$ such that each commit's creation time is no earlier than the issue's creation time and strictly earlier than one year after the issue's creation time. 
Next, as shown in lines 7 to 10 of the algorithm, we filter the $\mathit{candidate\_commits}$ to include all commits whose creation time is no earlier than the issue's creation time and strictly earlier than one year after the issue's creation time. 
Additionally, the commit should not be the fix commit of the issue. We then construct the $\mathit{false\_links}$ by pairing the $\mathit{issue}$ with each commit in $\mathit{candidate\_commits}$ and labeling the pair as 0. 
Finally, we append both the $\mathit{true\_links}$ and the $\mathit{false\_links}$ for the issue to the evaluation dataset and then proceed to process the next issue. 
Note that this algorithm is executed separately for each project, as each project has its own test dataset and commits pool. Following prior work~\cite{zhang2023ealink,ruan2019deeplink}, issue tags (e.g., ``\#123'', ``JIRA-456'') in the issue and commit text are removed to prevent potential data leakage.
% The constructed evaluation dataset for each project is presented in Table~\autoref{tab:merged_table}.

\begin{table}[t!]
\caption{
Statistics of the benchmark}
\label{tab:merged_table}
\small
\begin{tabular}{cccc}
\hline
Project name &\#Commits  & \#Unique issue\_id & \makecell{Average \# of \\ false links per issue} \\ \hline
Ambari$^\star$     & 24809& 1000 & 3978.26 \\
Calcite$^\star$     & 5889& 551  & 516.31  \\
Groovy$^\star$     & 20862& 1000 & 1064.79 \\
Ignite$^\star$     & 28869& 1000 & 2611.81 \\
Isis$^\star$       & 24945& 652  & 1265.91 \\
Netbeans$^\star$    & 10501 & 159& 1279.68 \\ \hline
Derby$^\dagger$       & 8040 & 43  & 363.93  \\
Drools$^\dagger$      & 16585 & 182& 630.74  \\
Infinispan$^\dagger$   & 17078& 399 & 1110.68 \\
Maven$^\dagger$        & 14685& 61 & 377.15  \\
Pig$^\dagger$         & 3675 & 45  & 197.44  \\ \hline
Flask$^\ddagger$     & 5353 & 151  & 357.88  \\
Keras$^\ddagger$      & 11248& 111 & 454.23  \\
Pgcli$^\ddagger$      & 2364 & 105 & 356.92  \\ \hline
Airflow$^\S$      & 27123 & 961& 1882.61 \\
Arrow$^\S$       & 16947 & 1000& 2058.99 \\
Beam$^\S$         & 43595& 865 & 6096.72 \\
Cassandra$^\S$    & 29867& 25  & 2560.84 \\
Flink$^\S$       & 35767& 1000 & 3251.68 \\
Freemarker$^\S$     & 2492 & 9 & 179.00  \\ \hline
\end{tabular}
\begin{minipage}{\linewidth}
\footnotesize
\textbf{Note:} Each project is annotated with a superscript representing its original dataset source: $^\star$ Zhang \etal{}~\cite{zhang2023ealink}, $^\dagger$ Dong \etal{}~\cite{dong2022semi}, $^\ddagger$ Lin \etal{}~\cite{lin2021traceability}, $^\S$ Mazrae \etal{}~\cite{mazrae2021automated}
\end{minipage}
\end{table}

~\autoref{tab:merged_table} shows the statistics of the evaluation dataset.
The dataset includes the same issues as used in the evaluation of prior studies~\cite{zhang2023ealink,dong2022semi,lin2021traceability,mazrae2021automated}.
Each issue has an average of 1,530 false links, with a minimum of 179 (Freemarker). 
% On average, each issue has 1,529.78 false links, with the Freemarker project having the lowest at 179.
In contrast, for each issue, EALink~\cite{zhang2023ealink} constructs a constant number of 99 false links per issue, which is substantially smaller than the number of commits within a one-year time frame from issue creation. 
Note that an issue may be fixed by one or more commits~\cite{zhang2023ealink}.
In our evaluation dataset, 17\% of issues were fixed by multiple commits.

%% file: 4_alg.tex
\SetKwFor{While}{while}{do}{}
\SetKwInput{KwInput}{Input}
\SetKwInput{KwResult}{Result}
\SetKw{KwReturn}{return}
\SetAlCapFnt{\footnotesize}
\SetKwComment{Comment}{$\triangleright$\ }{}
\SetCommentSty{mycommfont}
\DontPrintSemicolon

\begin{algorithm}[t!]
% \footnotesize
% \setstretch{1.05}
\caption{False Links Construction under RDS}
\label{alg:false_link_gen}

\KwInput{
    \textit{test\_set\_true\_links} \Comment*{ Contains true issue-commit pairs labeled as 1 }
    % \textit{commits\_pool}: Contains all available commits (Section~\autoref{sec:commits-pool})
}
\KwInput{
    % \textit{test\_set\_true\_links}: Contains issue-commit pairs labeled as 1 \;
    \textit{commits\_pool} \Comment*{Contains all fetched commits (\autoref{sec:commits-pool})}
}
\KwResult{
    \textit{evaluation\_dataset}
}

\SetKwProg{Fn}{Function}{:}{end}
\Fn{EvalSetGen(test\_set\_true\_links, commits\_pool)} {
    \textit{evaluation\_dataset} $\leftarrow$ $\emptyset$ \;
    \textit{sampled\_issue\_id\_list} $\leftarrow$ test\_set\_true\_links[\texttt{issue\_id}].unique()[:1000] \;

    \For{each \textit{issue\_id} in \textit{sampled\_issue\_id\_list}} {
        \textit{true\_links} $\leftarrow$ test\_set\_true\_links[issue\_id] \; 
        \textit{issue} $\leftarrow$ true\_links[0].issue\_info \; 
        
        \textit{candidate\_commits} $\gets$ \{ commit $\in$ commits\_pool $\mid$ \\
\hspace{1em} commit.time $\geq$ issue.create\_time \textbf{and} \\
\hspace{1em} commit.time $\leq$ issue.create\_time + $\epsilon$ \  \textbf{and} \\
% \hspace{1em} commit.id $\notin$ true\_links.commit\_id \} \;
\hspace{1em} commit $\neq$ issue.fix\_commit \} \;

        \textit{false\_links} $\leftarrow$ \{(issue, commit, label=0) $\mid$\\ 
     \hspace{1em}   commit $\in$ candidate\_commits \} \;

        \textit{evaluation\_dataset.append(false\_links)} \;
        \textit{evaluation\_dataset.append(true\_links)} \;
    }

    \KwRet \textit{evaluation\_dataset}
}
\end{algorithm}

%% file: 3_simple_method.tex
\section{The \tool{} Approach}
\label{sec:simple_method}

This section details \tool{}.~\autoref{fig:method} shows an overview of \tool{}, 
which consists of two stages: the first stage uses a vector database for scalable retrieval to fetch commits ranked by their similarity to the issue
% most similar commits, 
and the second stage prompts an LLM to rerank them by their relevance to the issue. 
We elaborate the key steps, vector database indexing and retrieval,
% (Section~\autoref{sec:faiss}), 
and LLM-assisted reranking below.
% (Section~\autoref{sec:llm_assist}).

% \subsection{Embedding Process}\label{sec:embedding_process}

\subsection{Vector Database Indexing and Retrieval}\label{sec:faiss}

The process begins by embedding the commit messages using an embedding tool, retaining the commit ID as metadata. Similarly, the issue summary and description are concatenated and embedded together, with the issue ID retained as metadata. To generate the embeddings, we utilize the Sentence-Transformers library~\cite{reimers2019sentence} with the \texttt{all-MiniLM-L6-v2} model, a lightweight and efficient transformer based on Microsoft's MiniLM architecture~\cite{wang2020minilm}. We chose this model because it is the most downloaded and most liked sentence-similarity model on Hugging Face, indicating strong community trust and widespread adoption. 
% We believe that other models with comparable performance would produce similar results, as discussed in Section~\autoref{sec:rq2}. 
This model employs self-attention distillation to capture contextual information effectively while maintaining computational efficiency. Given an input text \( T \), the embedding process can be defined as follows:
\begin{equation}
    \mathbf{E} = \text{MiniLM}(T)
\end{equation}
where \( \mathbf{E} \in \mathbb{R}^{384} \) represents the output embedding vector in a 384-dimensional space.

After generating embeddings, we use a vector database for indexing and retrieval. We use FAISS~\cite{douze2024faiss}—an off-the-shelf library designed for efficient nearest neighbor (NN) retrieval in high-dimensional spaces. We believe that switching FAISS with another vector database with comparable capability would produce similar results, as shown in ~\autoref{sec:rq3},

To retrieve commits from the vector database, we compute the cosine similarity score, following prior studies~\cite{zhang2023ealink,ruan2019deeplink,rath2018traceability}.
The computation is given by the following equation:
\begin{equation}
    \operatorname{sim}(\mathbf{q}, \mathbf{e}_i) \;=\;
    \frac{\mathbf{q} \cdot \mathbf{e}_i}{\lVert \mathbf{q}\rVert_2 \,\lVert \mathbf{e}_i\rVert_2}
\end{equation}
where $\mathbf{e}_i$ is the $i$-th element of the commit message embeddings
\(
\mathbf{E} = \langle \mathbf{e}_1, \mathbf{e}_2, \dots, \mathbf{e}_n \rangle
\)
that are indexed by the vector database for efficient access, and $\mathbf{q}$ is the issue embedding.
A higher score between the query issue and the commit message indicates greater similarity, enabling the ranking of commit messages and the generation of a list of candidate commits.

\subsection{LLM-Assisted Reranking}\label{sec:llm_assist}

Due to the semantic gap between issues and commits~\cite{zhang2023ealink, lin2021traceability, dong2022semi}, retrieving the most similar commit does not guarantee finding the one that fixed the issue.  
While the correct commit may have been retrieved, it could be obscured by incorrect commits that exhibit greater similarity to the issue.
To address this, we rerank the retrieved commits using a large language model (LLM).

In this phase, 
the top-$k$ commits from the initial retrieval are provided to the LLM with a structured prompt, as shown in~\autoref{fig:method}. 
The prompt includes the issue text (summary and description), along with each commit's ID and message. We also specify an output format instructing the LLM to return a reranked list of commit IDs. The LLM will analyze these commits and produce a reranked list based on contextual understanding and issue-commit relevance. 
Outputs that do not match the expected format (0.1\% of the time) default to the initial retrieval results.

In detail, we adopt a zero-shot approach guided by prompts with ChatGPT (\emph{gpt-4o}), following recent studies~\cite{wang2024enhancing,islam2024gpt,fuchss2025lissa}, which demonstrate \emph{gpt-4o} advanced capabilities in understanding complex textual relationships. 
The parameter $k$ controls the additional cost incurred for increasing precision. 
While it is possible to rerank all fetched commits for higher precision, this requires a longer prompt for the LLM, requiring more computation resources.
For the value of $k$, we select $k$=10 for its balance of precision and efficiency. We later show that this allows a high Precision@1 without incurring a high cost. 
This will be discussed in~\autoref{sec:rq3}.

%% file: 5_experimental_setup.tex
\section{Experimental Setup}
\label{sec:experimental_setup}

\subsection{Research Questions}

This work aims to answer the following research questions (RQs):

\vspace{0.2cm}\noindent{\bf RQ1: How does the state-of-the-art tool perform on a realistic evaluation dataset?  } 
This question investigates the performance of the state-of-the-art EALink~\cite{zhang2023ealink} on the evaluation dataset constructed under the Realistic Distribution Setting (RDS). 
First, we replicate the successful performance of EALink on their original evaluation dataset.
Next, we investigate the performance of EALink after expanding the evaluation onto a larger benchmark.
Afterwards, to understand the sensitivity of its performance to the evaluation dataset, we investigate how much the performance of EALink changes when evaluated under Realistic Distribution Setting \hhre{and compare it with the traditional IR method, VSM.} 
 
    \vspace{0.2cm}\noindent{\bf RQ2: How does \tool{} perform on the same realistic evaluation dataset?   } 
    This research question is concerned with the effectiveness of \tool{}. 
    We assess \tool{} in terms of both its ability to distinguish true links from false links and its efficiency.
    We analyze the sensitivity of the performance of \tool{} to the evaluation setup by comparing its performance on the different evaluation methods.
    
    \vspace{0.2cm}\noindent{\bf RQ3: Does \tool{}'s performance change under different configurations? } 
    \hhre{This question aims to investigate the effect of different settings of \tool{}. 
    We explore different embedding models for the vector database and various $k$ value settings for the LLM-assisted stage to assess whether these changes will significantly affect \tool{}'s performance.}

\subsection{Experiment Setting}

\subsubsection{Hardware Configuration}
The experiments were conducted on a machine equipped with two Intel(R) Xeon(R) Platinum 8480C CPUs @ 3.80GHz, 2.0 TiB of main memory, and one NVIDIA H100 80GB HBM3 GPU.

\subsubsection{LLM Setup}

We utilized the ChatGPT model GPT-4o provided by OpenAI, specifically the \texttt{gpt-4o-2024-11-20} version, with its default configuration. The temperature was set to the default value of 1.0, and the model was sampled once per query. The input token size was limited to a 128k token context window per API constraints, with no additional adjustments or fine-tuning.

\subsubsection{Baseline Description}
\label{sec:baseline_des}

We use EALink\footnote{We use EALink provided code and data from \url{https://github.com/KDEGroup/EALink}.}~\cite{zhang2023ealink} as the baseline, which is a state-of-the-art tool that outperforms T-BERT~\cite{lin2021traceability} and DeepLink~\cite{ruan2019deeplink}. It distills knowledge from CodeBERT~\cite{feng2020codebert} into a smaller model, fine-tuned with multi-task contrastive learning. 
We follow its original methodology, using the provided code and the same hyperparameters.
As a fundamental baseline, we also run VSM\footnote{We implemented VSM using the Gensim library (\url{https://pypi.org/project/gensim})}~\cite{salton1975vector}, which is widely used in other studies~\cite{zhang2023ealink,lin2021traceability,dong2022semi}.

\subsection{Evaluation Metrics}
\label{sec:eval_metrics}
We adopt the same metrics used by Zhang \etal{}~\cite{zhang2023ealink}, which use standard metrics for information retrieval tasks~\cite{radlinski2010comparing,sakai2008information}: 
Precision@$k$ (P@$k$), Normalized Discounted Cumulative Gain (NDCG@$k$), Mean Reciprocal Rank (MRR), and Hit Ratio (Hit@$k$) for evaluation. We also include Recall@$k$, which was overlooked in prior studies.

\begin{itemize}[leftmargin=*]

    \item \textbf{Precision@k} evaluates the proportion of relevant commits (\ie{} commits that belong to the correct issue-commit links) within the top \( k \) retrieved results:  
    \begin{equation}
        \text{Precision@k} = \frac{1}{|Q|} \sum_{i \in Q} \frac{\text{Rel}_i}{k},
    \end{equation}
    where \( Q \) is the query set, \( |Q| \) its size, and \( \text{Rel}_i \) the number of correctly linked commits in the top \( k \) results for query \( i \).

 \item \textbf{Hit@k} measures the likelihood that at least one correct commit appears within the top \( k \) retrieved results:  
    \begin{equation}
        \text{Hit@k} = \frac{1}{|Q|} \sum_{i} \mathbb{I} (\text{Rank}_i \leq k),
    \end{equation}
    where \( \mathbb{I}(\cdot) \) returns 1 if the highest-ranked relevant commit for query \( i \) is within the top \( k \), and 0 otherwise. Hit@1 is equivalent to Precision@1.

\item \textbf{Recall@k} evaluates the proportion of relevant commits retrieved within the top \( k \) results:  
    \begin{equation}
        \text{Recall@k} = \frac{1}{|Q|} \sum_{i \in Q} \frac{\text{Rel}_i}{\text{TotalRel}_i},
    \end{equation}
    where \( Q \) is the query set, \( |Q| \) its size, \( \text{Rel}_i \) the number of retrieved relevant commits, and \( \text{TotalRel}_i \) the total relevant commits for query \( i \).

     \item \textbf{MRR} (Mean Reciprocal Rank) evaluates how early the first relevant commit appears in the ranked list for each query:  
    \begin{equation}
        \text{MRR} = \frac{1}{|Q|} \sum_{i=1}^{|Q|} \frac{1}{\text{Rank}_i},
    \end{equation}
    where \( |Q| \) is the number of queries, and \( \text{Rank}_i \) is the position of the first correctly linked commit for query \( i \). A higher MRR indicates earlier retrieval of relevant commits.

    \item \textbf{NDCG@k} (Normalized Discounted Cumulative Gain) assesses how well relevant commits are ranked within the top \( k \) retrieved results:  
    \begin{equation}
        \text{NDCG@k} = \frac{1}{Z_k} \sum_{i=1}^{k} \frac{2^{r_i} - 1}{\log_2 (i + 1)},
    \end{equation}
    where \( Z_k \) is a normalization factor ensuring the ideal ranking achieves a value of 1. The term \( r_i \) represents the relevance score of the commit at position \( i \) (\( r_i = 1 \) for a correct match, 0 otherwise).

\end{itemize}

%% file: 5_results.tex
\section{Results}
\label{sec:results}

\subsection{RQ1: How does the state-of-the-art tool perform on a realistic evaluation dataset?}
\label{sec:rq1}

To ensure we replicate EALink~\cite{zhang2023ealink} correctly and perform a fair comparison, we ran EALink on both the evaluation dataset constructed using its original method and the dataset created under the Realistic Distribution Setting (RDS), then compared the results.
% Note that the RDS method was only used to produce the evaluation dataset. 
In the original experiments of Zhang \etal{}, EALink was trained using a balanced dataset and then tested on an imbalanced dataset. We reused the same code for training. 

\begin{table}[t!]
\centering
\caption{\label{tab:results_ealink_drop}Performance of EALink on evaluation datasets constructed using different methods}
\begin{tabular}{ccc}
\hline
            & False Links=99 & RDS   \\ \hline
P@1 (Hit@1) & 32.52        & 14.43 (\(\downarrow 55.63\%\))   \\
P@10        & 7.08         & 3.64 (\(\downarrow 48.59\%\))    \\
Hit@10      & 59.46        & 30.76 (\(\downarrow 48.27\%\))   \\
Recall@10      & 56.68        & 28.06 (\(\downarrow 50.49\%\))   \\
MRR         & 41.16        & 20.21 (\(\downarrow 50.90\%\))   \\
NDCG@1      & 20.52        & 9.10 (\(\downarrow 55.65\%\))   \\
NDCG@10     & 30.84        & 15.52 (\(\downarrow 49.67\%\))   \\ \hline
\end{tabular}
\begin{minipage}{\linewidth}
\footnotesize
\textbf{Note:} Results are averaged over 20 projects. The column False Links=99 shows EALink's performance on a dataset constructed with a fixed 99 false links per issue, while the column RDS shows its performance on a dataset constructed under Realistic Distribution Setting (RDS). The parentheses in the RDS column show the percentage change in performance. 
\end{minipage}
\end{table}

Following EALink's original evaluation~\cite{zhang2023ealink}, we first used their false link generation method to construct an evaluation dataset using projects provided by them: Ambari, Calcite, Groovy, Ignite, Isis, and NetBeans.
% from the evaluation subjects: Ambari, Calcite, Groovy, Ignite, Isis, and NetBeans.
For any given issue, it randomly samples 99 of the commits in the ground truth test dataset to construct false links. 
% In our case, we found that in some projects, the test set was too small to contain 99 commits, so we used all remaining commits as candidates. 
Next, with this evaluation dataset, which we refer to as the original evaluation dataset, 
we evaluated EALink. 
EALink obtains an average Precision@1 of 53.67\%, which matches the 53.90\% Precision@1 reported in the paper, indicating that our replication was successful. 
Then, we expanded the evaluation to 20 projects (introduced in ~\autoref{sec:ground-truth}). 
The results, shown in the first column of~\autoref{tab:results_ealink_drop},
indicate an average Precision@1 of 32.52\%. 

Then, we ran EALink on the evaluation dataset constructed under 
% RDS %Adaptiv Link Generation
the Realistic Distribution Setting. 
Unlike the original evaluation dataset of Zhang \etal{}~\cite{zhang2023ealink}, which had a fixed number of 99 false links per issue, our evaluation  
% adapts based on repository commit activity, 
generates more false links for repositories with a larger number of commits. 
This results in an average number of false links per issue of 1,530. 
In the second column of~\autoref{tab:results_ealink_drop}, we present the results of running the tool on our evaluation dataset, which shows a significant performance drop compared to the original evaluation dataset. 
Across the 20 projects, EALink's Precision@1 decreases to 14.43\%, a decline of 55.63\%. Similarly, Hit@10 drops from nearly 60\% to about 30\%, reducing by half. 
On the six projects provided by EALink (including Ambari, Calcite, and four others), the average Precision@1 decreases from 53.67\% to 28.05\%.
Additionally, we evaluated the traditional IR method, VSM, on the same evaluation dataset.
Its results are presented in~\autoref{tab:results_compare_all}. 
Surprisingly, EALink underperforms the traditional IR method, VSM. 
VSM achieves a Precision@1 of 46.59\% and demonstrates better performance across all other metrics.
These results suggest that the use of VSM would be preferred over EALink in a realistic setting.

\begin{center} % 居中放置
{\setlength{\fboxsep}{6pt}
\colorbox{lightergray}{%
  \parbox{0.95\linewidth}{%
    \textbf{Answer to RQ1: } 
We constructed an evaluation dataset under the Realistic Distribution Setting (RDS), 
which considers more false links per issue when there is higher commit activity during a given period, better reflecting the practical use of an issue-commit link recovery technique on a repository. 
Under this evaluation, the average Precision@1 of the state-of-the-art tool declines to 14.43\% compared to a VSM baseline with a Precision@1 of 46.59\%.
  }%
}}
\end{center}

\subsection{\hhre{RQ2: How does \tool{} perform on the same realistic evaluation dataset?  }}

\label{sec:rq2}

\subsubsection{Comparison of Effectiveness}
\label{sec:results_compare}

\begin{table}[t!]

\caption{\label{tab:results_on_original_dataset} Comparison of linking effectiveness on the original dataset with a constant 99 false links constructed per issue. This comprises the six projects provided by Zhang \etal~\cite{zhang2023ealink} (Ambari, Calcite, Groovy, Ignite, Isis, and NetBeans)}
% \begin{threeparttable}
\begin{tabular}{cccc}
\hline
 & EALink & Vector DB & \tool{} \\ \hline
P@1   (Hit@1) & 53.67 & 81.83 & \textbf{90.04}(\(\uparrow 67.77\%\)) \\
P@10 & 8.94 & \textbf{11.14} & \textbf{11.14}(\(\uparrow 24.61\%\)) \\
Hit@10 & 72.56 & \textbf{92.89} & \textbf{92.89}(\(\uparrow 28.02\%\)) \\
Recall@10 & 69.03 & \textbf{89.93} & \textbf{89.93}(\(\uparrow 30.28\%\)) \\
MRR & 60.13 & 85.87 & \textbf{91.38}(\(\uparrow 51.97\%\)) \\
NDCG@1 & 33.86 & 51.63 & \textbf{56.81}(\(\uparrow 67.78\%\)) \\
NDCG@10 & 41.33 & 56.31 & \textbf{58.07}(\(\uparrow 40.50\%\)) \\ \hline
\end{tabular}
\begin{minipage}{\linewidth}
\footnotesize
\textbf{Note:} The parentheses in the \tool{} column show the improvements over EALink. Bold numbers indicate the highest performance. All values are in \%.
\end{minipage}
% \end{threeparttable}
\end{table}

\begin{table}[t!]
\centering
% \small
\setlength{\tabcolsep}{4pt}

\caption{\label{tab:results_compare_all} Comparison of linking effectiveness on the dataset constructed under RDS}

% \begin{threeparttable}
\begin{tabular}{ccccc}
\hline
Metric      & EALink  &VSM& Vector DB      & \tool{}  \\ \hline
P@1 (Hit@1) & 14.43 &46.59  & 61.57  & \textbf{75.03} (\(\uparrow 420.0\%\)) \\
P@10        & 3.64 & 8.70  & \textbf{10.09}  & \textbf{10.09} (\(\uparrow 177.2\%\)) \\
Hit@10      & 30.76 &70.94  & \textbf{83.38}  & \textbf{83.38} (\(\uparrow 171.1\%\)) \\
Recall@10      & 28.06 & 67.22 & \textbf{78.84}  & \textbf{78.84} (\(\uparrow 181.0\%\)) \\
MRR         & 20.21 & 55.10  & 69.21  & \textbf{78.52} (\(\uparrow 288.5\%\)) \\
NDCG@1      & 9.10  &29.39  & 38.85  & \textbf{46.86} (\(\uparrow 415.0\%\)) \\
NDCG@10     & 15.52 & 39.45 & 47.98  & \textbf{51.08} (\(\uparrow 229.1\%\)) \\ \hline
\end{tabular}

\begin{minipage}{\linewidth}
\footnotesize
\textbf{Note:} Results are averaged over 20 projects (\%). The parentheses show the improvements over EALink. Since P@10, Hit@10, and Recall@10 only assess whether the true commits are included within the top 10 results, without considering its exact rank, they are the same for both the vector database and \tool{}. Bold numbers indicate the highest performance. 
\end{minipage}
% \end{threeparttable}

\end{table}

\hhre{To ensure a fair comparison, we first evaluated the vector database and \tool{} on the dataset constructed using the EALink method~\cite{zhang2023ealink}, which generates 99 false links per issue and includes the six projects provided by EALink. The results, shown in~\autoref{tab:results_on_original_dataset}, indicate that \tool{} achieves a Precision@1 of 90.04\%, significantly outperforming EALink’s 53.67\%. The vector database method also outperforms EALink, achieving a Precision@1 of 81.83\%.}

Furthermore, we compared them using our more realistic evaluation dataset. As presented in~\autoref{tab:results_compare_all}, the vector database approach achieves an average Precision@1 of 61.57\%, a significant improvement over EALink’s 14.43\%. With LLM-assisted reranking, \tool{} further enhances Precision@1 by an average of 13.46\%, reaching 75.03\%,  representing a 420.0\% increase compared to EALink. Following the Mann-Whitney U test~\cite{mann1947test}, the improvement of \tool{} over EALink in every metric is statistically significant (p-value $<$ 0.01) and exhibits a large effect size~\cite{cohen2013statistical} (Cohen’s D $>$ 0.8).
These results demonstrate that \tool{} is effective in issue-commit link recovery and that LLMs can be effectively leveraged to enhance performance.

\subsubsection{Comparison of Efficiency}
\label{sec:vectordb_ealink_efficiency}

\begin{table}[t!]
    \centering

    \caption{\label{tab:time_cost}Comparison of training and testing time cost }

    \small
    % \normalsize
    \begin{tabular}{cccc}
        \hline
         & EALink &  Vector DB &  \tool{} \\ \hline
        \makecell{Train\\(Total / Per Link)} &  100.68h / 4.39s &  N/A &  N/A \\
        \makecell{Test\\(Total / Per Issue)}  & 17.78h / 6.87s  & 2.26h / 0.87s & 14.02h / 5.42s \\ \hline
    \end{tabular}
\begin{minipage}{\linewidth}
\footnotesize
\textbf{Note:} ``Total'' indicates the overall time for all 20 projects. ``Per Link'' (training) is the average time per link, computed as total training time divided by the number of training links. ``Per Issue'' (testing) is the average time per issue, based on total testing time divided by the number of evaluation issues. Testing times for Vector DB and \tool{} include embedding, indexing, and retrieval.
\end{minipage}
    
\end{table}

Efficiency is an important aspect of software traceability, especially when used in large-scale industrial settings~\cite{cleland2012software,spanoudakis2005software}.
Therefore, we calculated the training and testing times required for three approaches—EALink, the vector database, and the \tool{}—to compare their efficiency. The results, shown in~\autoref{tab:time_cost}, indicate that for the 20 projects, EALink requires 100.68 hours for training, with an average of 4.39 seconds per link in the training set. However, approaches using a vector database do not require training on a specific issue-commit link dataset, making them easier to use and more practical with less human effort.

Regarding testing time, EALink requires a total of 17.78 hours for the 20 projects, averaging 6.87 seconds per issue, while the vector database approach takes only 2.26 hours in total, averaging 0.87 seconds per issue.
These differences are statistically significant according to the Mann-Whitney U test~\cite{mann1947test} (p-value $<$ 0.01) and demonstrate a large effect size (Cohen’s D $>$ 0.8).
Additionally, the LLM-assisted method (\ie{} \tool{}) takes 14.02 hours for testing, which is faster than EALink and achieves a significantly higher performance. 
\hhre{These results demonstrate that the use of vector database is an efficient solution.} 
It eliminates the need for pretraining models and reduces testing time. 
By incorporating an LLM reranking step, \tool{} achieves a high precision while eliminating the need for training models. In practice, each issue requires about 758 input tokens and 266 output tokens, leading to an estimated cost of only \$0.009 per issue when using GPT-4o.

\begin{center} % 居中放置
{\setlength{\fboxsep}{6pt}
\colorbox{lightergray}{%
  \parbox{0.95\linewidth}{%
    \textbf{Answer to RQ2:} 
Our results show that on the realistic evaluation dataset, \tool{} achieves a Precision@1 of 75.03\%, outperforming EALink (which achieves 14.43\%) by 420.0\%. Moreover, using the vector database-based method eliminates the need for pretraining on task-specific data, making it a more efficient solution.
  }%
}}
\end{center}

\subsection{\hhre{RQ3: Does \tool{}'s performance change under different configurations?}}
\label{sec:rq3}

\subsubsection{\hhre{Vector Database: Evaluating Different Embedding Models}}
\label{sec:diff_emb}

We evaluated two additional embedding models for use in the vector database. The first model, \texttt{all-mpnet-base-v2}, built upon Microsoft's MPNet architecture~\cite{song2020mpnet}, is a sentence transformer and the second-most-downloaded sentence-similarity model on Hugging Face (the most popular is 
% the one we used
\hhre{used in \tool{}}). The second model, \texttt{text-embedding-ada-002}, is the default model for OpenAI embeddings~\cite{neelakantan2022text}, which is a larger 
% cloud-based 
model designed for higher precision. 

\begin{table}[t!]
\caption{\label{tab:diff_emb} Comparison of linking effectiveness while varying the choice of embedding method}
\begin{tabular}{cccc}
\hline
 & MiniLM & MPNet & OpenAI \\ \hline
P@1 (Hit@1) & 61.57 & 62.73 & 70.10 \\  
P@10 & 10.09 & 10.18 &  10.62\\ 
Hit@10 & 83.38 & 83.00 & 87.08 \\  
Recall@10 & 78.84 & 78.40 &  82.91\\  
MRR & 69.21 & 70.39 &  76.51\\  
NDCG@1 & 38.85 & 39.58 &  44.22\\  
NDCG@10 & 47.98 & 48.32 & 51.56 \\  
Test Time (hour) & 2.26h & 3.22h & 28.72h \\
Model Size & 22.7M params & 109M params & Cloud-based \\ \hline
\end{tabular}
\begin{minipage}{\linewidth}
\footnotesize
\textbf{Note:} Results are averaged over 20 projects. All values are in \%, except for test time and model size.
\end{minipage}
\end{table}

~\autoref{tab:diff_emb} presents the results comparing 
% these two embedding models with the one  we used
\hhre{three embedding models}. The comparison between MiniLM 
% our chosen model
(\hhre{used in \tool{}}) and MPNet shows that 
the model we selected in our tool requires less time while achieving comparable performance. Although OpenAI embeddings provide better performance—with Precision@1 increasing from 61.57\% to 70.10\% (an improvement of 8.53\%)—they require 28.72 hours to complete the experiment, approximately 11 times longer than the MiniLM model. Given the high computational cost, MiniLM is a more practical option.

\subsubsection{\hhre{LLM-Assisted Reranking: Exploring Different Top-$k$ Values}}
\label{sec:diff_top_k}

To further analyze the effect of the $k$ setting, we conduct experiments with different top-$k$ values, specifically $k = 5, 10, 15, 20$. 
% This evaluation helps us understand how different numbers of candidates for reranking affect overall performance and time cost. 
\hhre{We conducted this experiment across 20 projects using the realistic evaluation dataset.} 
The results in~\autoref{fig:reranking_topk} show that increasing $k$—meaning reranking a larger set of top results—generally leads to better performance, specifically higher Precision@1 and NDCG@1. However, this improvement comes with a significant increase in time cost and requires processing more input tokens, leading to a higher cost in invoking OpenAI API. While setting $k=5$ results in relatively lower performance, increasing $k$ to 10, 15, or 20 does not lead to significant performance gains. Considering performance, time, and resource costs, we find that $k=10$ provides the best trade-off. Additionally, compared to the straight line in the figure representing the Precision@1 of EALink's when run on the same evaluation dataset, our approach outperforms EALink across different $k$ values. 
This shows that its performance does not rely on tuning $k$.

\begin{center} % 居中放置
{\setlength{\fboxsep}{6pt}
\colorbox{lightergray}{%
  \parbox{0.95\linewidth}{%
    \textbf{Answer to RQ3:} 
For vector database retrieval, using different embedding models with similar runtime costs yields comparable results. In the LLM-assisted reranking stage, increasing the $k$ value consistently improves performance over the baseline but increases test time cost.
  }%
}}
\end{center}

%% file: 6_discussion.tex
\section{Discussion}
\label{sec:discussion}

\subsection{Revisiting Classical Baselines}
In addition to recent learning-based methods, it is important to include classical baselines when revisiting research progress on issue-commit link recovery. We therefore evaluate ReLink~\cite{wu2011relink}, an approach that extends beyond the basic VSM, using our constructed dataset. Since ReLink outputs positive issue-commit links rather than ranked candidates, we aligned the evaluation by treating \tool{}’s top-ranked commit per issue as its predicted link and computed precision, recall, and F1 score accordingly. The average results across 20 projects show that ReLink achieves 10.92\% precision, 12.59\% recall, and 10.79\% F1 score, whereas \tool{} reaches 75.03\% precision, 50.75\% recall, and 58.75\% F1 score, significantly outperforming ReLink. Unlike ReLink, which relies on explicit issue tags in commit messages (removed in our setup) and operates at the set level, \tool{} leverages semantic relevance and ranks candidates per issue, making it more robust in large and noisy candidate pools.

\subsection{Lessons learned and Implications}

\vspace{0.2cm}
\noindent{\bf Keep your feet on the ground -- evaluation should match practice.}
% Issue-commit linking recovery is an important task in software development, and its
Our results suggest that the performance of issue-commit linking approaches is sensitive to their evaluation setups, which emphasizes the importance of the evaluation of proposed techniques to reflect the conditions under real-world practices. 
Future research should use evaluations that better reflect a practical usage scenario.

\begin{figure}[t!]
    \centering
    \includegraphics[width=0.49\textwidth]{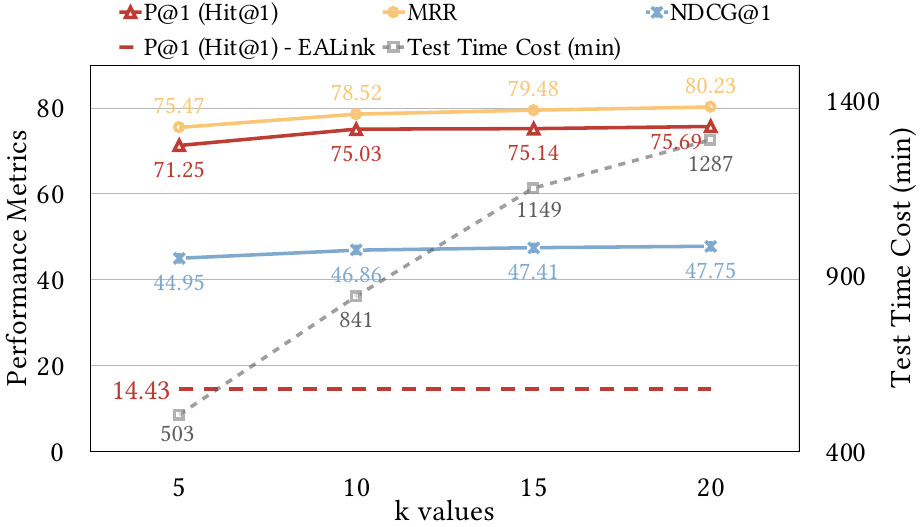}
    \caption{Effect of varying $k$ in \tool{} during reranking: A higher $k$ slightly raises performance (left axis: P@1, MRR, NDCG@1) while significantly increasing test time cost (right axis).
    }
    \label{fig:reranking_topk}
\end{figure}

\vspace{0.2cm}
\noindent{\bf Back to the basics -- IR techniques are strong baselines.} 
Our results showed that the VSM baseline, a dated IR model~\cite{antoniol2002recovering} proposed by Wu \etal{}~\cite{wu2011relink} for issue-commit linking, was effective. 
This is consistent with the ``Easy over hard'' principle advocated by Fu and Menzies~\cite{fu2017easy}.
This finding has practical implications -- practitioners may find that the use of simple and fast approaches match the performance of more complex approaches.

\vspace{0.2cm}
\noindent{\bf Don't forget your roots -- updating baselines.}

Studies on issue-commit linking have included traditional IR baselines, such as the Vector Space Model (VSM)~\cite{salton1975vector}, Latent Dirichlet Allocation (LDA)~\cite{blei2003latent}, and Latent Semantic Indexing (LSI)~\cite{deerwester1990indexing}.
However, recent studies continue to rely on traditional methods as baselines without considering recent advances in the IR literature. 
In our experiments, the out-of-the-box use of a modern baseline, the vector database FAISS~\cite{douze2024faiss}, already achieves a strong performance.
This shows that vector databases provide an effective retrieval approach for issue-commit linking, highlighting the need to update baselines to reflect the latest advancements in IR methods. 
We call for the need for newly proposed methods to be evaluated with baselines that are continuously updated to accurately measure real progress.

% \textbf{Effectiveness of LLM for Reranking}  
\vspace{0.2cm}
\noindent{\bf From good to great – effectiveness of LLMs in retrieval refinement.}
Our experiments demonstrate that a reranking step using LLMs %large language models (LLMs) 
is highly effective in overcoming the semantic gap, consistent with findings in information retrieval systems~\cite{zhu2023large,sun2023chatgpt,gao2024llm}. To assess whether LLM-based reranking can also enhance the performance of EALink~\cite{zhang2023ealink}, a deep learning-based method, we conducted experiments on our realistic evaluation dataset, including 20 projects. As shown in~\autoref{tab:ealink_llm}, with LLM reranking assistance, we observe a Precision@1 improvement of 80.45\%, increasing from 14.43\% to 26.04\%. This demonstrates the LLM’s effectiveness in refining initially imprecise results.
Future work can explore the optimization of LLM prompts and apply domain-specific fine-tuning to further improve the refinement step.

\begin{table}[t!]
\centering
\caption{Comparison of EALink and EALink+LLM}
\label{tab:ealink_llm}
\renewcommand{\arraystretch}{1.2}
\begin{tabular}{cccc}
\hline
Metric       & EALink  & EALink+LLM  & Improvement  \\ \hline
P@1 (Hit@1)  & 14.43   & 26.04       & \textbf{$\uparrow$ 80.45}  \\  
P@10         & 3.64    & 3.64       & -                 \\  
Hit@10       & 30.76   & 30.76      & -                 \\  
Recall@10       & 28.06   & 28.06      & -                 \\  
MRR          & 20.21   & 28.55       & \textbf{$\uparrow$ 41.26}  \\  
NDCG@1       & 9.10    & 16.96       & \textbf{$\uparrow$ 86.37}  \\  
NDCG@10      & 15.52   & 18.79       & \textbf{$\uparrow$ 21.07}  \\  
\hline
\end{tabular}

\begin{minipage}{\linewidth}
\footnotesize
\textbf{Note:} Results (\%) are averaged over 20 projects on our dataset constructed under the Realistic Distribution Setting (RDS).
 
\end{minipage}
\end{table}

%% file: 8_threats.tex
\subsection{Threats to Validity}
\label{sec:threats}

\vspace{0.2cm}
\noindent{\bf Threats to Internal Validity.} Threats to internal validity refer to errors in our experiments or implementation issues. 
To avoid implementation errors, we replicated the baseline tool, EALink~\cite{zhang2023ealink}, using its publicly available code. 
We ensured that we replicated its previously reported results before extending the experiments. 
Therefore, the threats to internal validity are minimal.

\vspace{0.2cm}
\noindent{\bf Threats to Construct Validity.} A potential threat to construct validity is the selection of evaluation metrics. 
We use widely adopted metrics—Precision@$k$, Hit@$k$, MRR, and NDCG@$k$—from prior studies~\cite{zhang2023ealink,dong2022semi,ruan2019deeplink,mazrae2021automated} and information retrieval tasks~\cite{radlinski2010comparing,sakai2008information}. 
We also included Recall@k, a standard information retrieval evaluation metric, overlooked in prior issue-commit linking work.
Consequently, we believe that any threat to construct validity is minimal.

\vspace{0.2cm}
\noindent{\bf Threats to External Validity.} Threats to external validity refer to factors that might limit the generalizability of our findings.
% In this context, the output consistency of large language models (LLMs) become a key consideration. 
% To address this, 
Our benchmark is the largest in the literature, which provides confidence that our findings are not specific to only a few projects.
One threat is the number of times our experiments were performed. 
% We evaluated the stability of our tool by running it multiple times on the Ignite evaluation dataset. 
\hhre{To mitigate the effect of randomness, we repeated our experiments five times. 
For all metrics, the average results exhibit standard deviations below 1\%.
} 
Given the stability of the results, repeating the experiments would not yield different findings. 
As such, there are minimal threats to external validity.
% Consequently, we believe that our work can be easily replicated by others.

%% file: 7_related_work.tex
\section{Related Work}

\label{sec:related_work}

\vspace{0.2cm}
\noindent{\bf Traceability Link Recovery:} 
Traceability link recovery methods create links between artifacts such as requirements, design documents, architecture models, and source code. 
Research has applied classic IR techniques~\cite{borg2014recovering,lucia2007recovering,gethers2011integrating,oliveto2010equivalence,eaddy2008cerberus,9678546}. 
Recent work~\cite{rodriguez2023prompts,fuchss2025lissa} utilized  LLMs but also found that their level of effectiveness has still been unable to support practical automatic link recovery~\cite{fuchss2025lissa,hayes2006advancing}.

\vspace{0.2cm}
\noindent{\bf Traditional 
% Feature- and Rule-Based 
Approaches for Issue-Commit Linking: } 
Traditional approaches combine heuristics and expert annotation to link commits with bug reports. Bachmann \etal{}~\cite{bachmann2010missing} used interactive heuristic linking. Wu \etal{}~\cite{wu2011relink} filtered candidates by textual similarity, timing, and committer mapping. It also learned optimal thresholds from training data. Nguyen \etal{}~\cite{nguyen2012multi} improved performance by adding code change analysis. Schermann \etal{}~\cite{schermann2015discovering} leveraged developer identity, time proximity, and resource overlap.
These methods often miss links~\cite{bird2009fair}.

\vspace{0.2cm}
\noindent{\bf Machine Learning-Based Approaches: } 
Machine learning methods improve linking accuracy by reducing the need for handcrafted heuristics. 
Le \etal{}~\cite{le2015rclinker} enriched commit messages via code summarization. 
Sun \etal{}~\cite{sun2017frlink} refined feature extraction with non-source documents and code file filtering. Sun \etal{}~\cite{sun2017improving} used positive-unlabeled learning to address limited labeled data. 
Rath \etal{}~\cite{rath2018traceability} combined process, stakeholder, structural, and textual similarity metrics. 
Mazrae \etal{}~\cite{mazrae2021automated} incorporated non-textual cues such as authorship, timestamps, and status. Dong \etal{}~\cite{dong2022semi}, a semi-supervised framework, tackled data imbalance and sparsity.
These approaches were later improved by deep learning.

\vspace{0.2cm}
\noindent{\bf Deep Learning-Based Approaches: } 
Recent works focused on the use of deep learning.
Ruan \etal{}~\cite{ruan2019deeplink} learned semantic representations with word embeddings and RNNs, while Xie \etal{}~\cite{xie2019deeplink} combined RNNs with SVMs and a code knowledge graph from ASTs for semantic and code context. Lin \etal{}~\cite{lin2021traceability} leveraged a BERT-based framework pre-trained on CodeSearchNet and fine-tuned on small datasets to address data sparsity. Zhu \etal{}~\cite{zhu2024deep} employed deep semi-supervised learning and iteratively retrained its model using pseudo-labels on unlabeled data. Zhang \etal{}~\cite{zhang2023ealink} 
employed knowledge distillation to improve both accuracy and efficiency.
% built on CodeBERT~\cite{feng2020codebert} and 
% and multi-task contrastive learning and outperformed previously proposed approaches.  
Despite these improvements, we found that the approaches were assessed using evaluations whose realism could be improved.
 
 % \tool{} mitigates these limitations by reranking commits using an LLM.

\vspace{0.2cm}
\noindent{\bf Replication Studies: } 
Our study found that the data used in evaluations for issue-commit link recovery may have the drawback of an unrealistic distribution of false links.
Some other studies~\cite{shepperd2013data,ding2024vulnerability,chakraborty2024revisiting,liu2025automatically,kang2022detecting,10.1145/3708525} have also emphasized the importance of using methods and data that evaluate tools in different settings. 
While some works~\cite{dong2022semi,lin2021traceability,mazrae2021automated,ruan2019deeplink,zhang2023ealink} have investigated issues of data cleanliness and data leakage, our work is the first replication study of issue-commit linkers and highlights the importance of evaluating them using realistic evaluation data that match the real development history.

%% file: 9_conclusion.tex
\section{Conclusion and Future Work}
\label{sec:conclusion}

In this study, we successfully replicate the strong performance of the state-of-the-art work on issue-commit linking. 
To investigate it further, we constructed a new benchmark under a proposed Realistic Distribution Setting (RDS) that adaptively constructs false links based on the level of development activity in the same time frame as the ground-truth link.
To the best of our knowledge, the benchmark is the largest in the literature, consisting of 9,319 unique issues from 20 open-source projects, with an average of 1,530 false links constructed per issue.
% While prior studies have considered canonical Information Retrieval-based baselines, we find that prior studies have overlooked the inclusion of a modern Information Retrieval-based baseline. 
We find that the use of an off-the-shelf IR method 
% \zt{do you mean ``vector'' database?}\hh{i think it is vsm, the traditional ir method.}  
outperforms the state-of-the-art technique.
Building on our findings, we propose \textbf{\tool{}}, a scalable and efficient approach that combines retrieval with an additional step of reranking using an LLM. 
% \tool{} outperform state-of-the-art methods.  
In terms of average Precision@1, \tool{} 
outperforms the state-of-the-art approach by more than four times.
% achieves an average Precision@1 of 75.04\%, compared to 14.43\% for the state-of-the-art, 
% validating the effectiveness of our approach.

In the future, we plan to extend \tool{} for other tasks in software traceability, such as mapping features to their implementations and linking requirements to code changes.

%% file: acknowledgement.tex
\begin{acks}
This research / project is supported by the National Research Foundation, Singapore, and the Smart Nation Group under the Smart Nation Group's Translational R\&D Grant (Award No. TRANS2023-TGC02). Any opinions, findings and conclusions or recommendations expressed in this material are those of the author(s) and do not reflect the views of National Research Foundation, Singapore or the Smart Nation Group.
\end{acks}